\documentclass[epj]{svjour}
\usepackage{amsmath,amssymb,graphicx,bm}
\usepackage{color}
\usepackage{amsmath,amssymb,bm,natbib}
\usepackage{epsfig}
\usepackage{graphicx}
\usepackage{slashed}
\usepackage{multirow}
\newcommand{\beq}{\begin{equation}}
\newcommand{\eeq}{\end{equation}}
\newcommand{\bea}{\begin{eqnarray}}
\newcommand{\eea}{\end{eqnarray}}
\newcommand{\ba}{\begin{align}}
\newcommand{\ea}{\end{align}}
\newcommand{\bfig}{\begin{figure}}
\newcommand{\efig}{\end{figure}}

\newcommand{\D}{\displaystyle}

\newcommand{\gev}{\, \text{GeV}}

\newcommand{\tin}{t_{\rm in}}
\newcommand{\la}{\langle}
\newcommand{\ra}{\rangle}

\newcommand{\omnes}{{\cal{O}}}

\newcommand{\fmsq}{\, \text{fm}^2}

\begin{document}

\vspace{1cm}

\title{Parametrization-free determination of the shape parameters
for the pion electromagnetic form factor}
\author{B. Ananthanarayan$^{a}$\and I. Caprini$^{b}$ \and Diganta Das$^{c}$ 
\and I. Sentitemsu Imsong$^{d}$ }
\institute{$^a$ Centre for High Energy Physics,
Indian Institute of 
Science, Bangalore 560 012, India \\
$^b$ Horia Hulubei National Institute for Physics and Nuclear Engineering,
P.O.B.MG-6, 077125 Bucharest-Magurele, Romania\\
$^c$ Institute of Mathematical Sciences,
 Taramani, Chennai 600113, India\\
$^d$Theoretische Physik 1, Naturwissenschaftlich-Technische Fakult\"{a}t, 
Universit\"{a}t Siegen, D-57068 Siegen, Germany}

\abstract{Recent data from high statistics experiments that have measured
the modulus of the  pion electromagnetic form factor from threshold to relatively high
energies are used as input in a suitable mathematical framework of
analytic continuation to find stringent constraints on the shape
parameters of the form factor at $t=0$. The method uses also as
input a precise description of the phase of the form factor in the elastic region based on
 Fermi-Watson theorem and the analysis of the $\pi\pi$ scattering amplitude with dispersive 
Roy equations, and some information on
the spacelike region coming from recent high precision
experiments. Our analysis confirms the inconsistencies of several data on the modulus, 
especially from low energies, with analyticity and the input phase, noted in our earlier work.  
Using the data on the modulus from energies above  $0.65 \,{\rm GeV}$,
 we obtain, with no specific parametrization, the prediction $\langle r_\pi^2 \rangle \in\, (0.42,\,0.44)\,\fmsq $ 
for the charge radius.
The same formalism leads also to very narrow allowed ranges for the higher-order 
shape parameters  at $t=0$, with a strong correlation among them.}


\titlerunning{Shape parameters of the pion form factor}

\authorrunning{B. Ananthanarayan et al.} 
\maketitle
\section{Introduction}
\label{sec:intro}

The pion electromagnetic form factor is an important 
quantity that encodes information on the structure of the strong
interactions.  At high spacelike momenta, it acts as an excellent
testing ground for studying the onset of perturbative QCD for exclusive quantities, while
at low energies it is an important laboratory for the study of chiral symmetry breaking. It plays a major role in the precision tests of  the Standard
Model, through its contribution  to the anomalous magnetic moment of the muon, which is one
of the most precisely measured observables in particle physics.

 The derivatives of the form factor at $t=0$ present in the Taylor series expansion 
\begin{equation}\label{eq:taylor}
	F(t) = 1 + \D\frac{1}{6} \la r^2_\pi \ra t + c t^2 + d t^3 + \cdots
\end{equation}
 are quantities of interest for  testing the expansions of chiral perturbation theory (ChPT) and the lattice calculations of the pion form factor. 
These shape parameters, especially the charge radius squared $\la r^2_\pi \ra$,  are the object of many investigations. 
However, the point $t=0$ is not directly accessible to experiment.
  Measurements near the
origin \cite{Amendolia:1986wj}-\cite{Huber} in the spacelike region $t<0$  have been exploited in the past 
 to yield the charge radius and the higher shape parameters.
Recent high statistics experiments 
\cite{BABAR}-\cite{Fujikawa:2008ma}  have measured the modulus in the timelike region on the unitarity cut.
Since this experimental information is available only quite far away from the origin, suitable techniques of extrapolation have to be invented in order to obtain a
reliable extraction of the shape parameters from these data.  

The analyticity properties of the form factor  are very useful  in performing this extrapolation. From causality and unitarity it is known that $F(t)$ is an analytic function in the complex $t$-plane, with a cut starting at the unitarity threshold $t_+= 4M_\pi^2$
and running to infinity. The Fermi-Watson theorem relates the
phase of the form factor in the elastic region,  $t_+ \le t\le \tin$, where $\tin$ is the first inelastic threshold, to the  $P$-wave phase shift of $\pi\pi$ scattering, which is now known to high
precision due to studies based on Roy equations for the $\pi\pi$ elastic amplitude
\cite{ACGL}-\cite{GarciaMartin:2011cn}.

Analyticity was exploited  in the past by means of various types of dispersive representations, which express the function in terms of either the imaginary part, the phase or the modulus on the cut. However, none of the standard dispersion relations has complete input, 
and their application requires {\it ad-hoc} assumptions.  Specific parametrizations of the data were also frequently used,  
but they are affected by the known ``instability" of analytic continuation \cite{Ciulli} if they are extrapolated  outside their original range of validity.

 Recently, more sophisticated methods of complex analysis were applied in a formalism  known as ``theory
of unitarity bounds" (for a review and earlier references see \cite{Abbas:2010EPJA}). 
The main feature of this approach is that it allows the optimal implementation of  the  phase and modulus information available on the unitarity cut.  Using mathematical techniques  belonging to the analytic interpolation theory on Hardy spaces, one can include as input also  values of the function or its derivatives at discrete points inside the analyticity domain.   This input is shown to lead to strong constrains on other values of the form factor, like the shape parameters at the origin  or the  values on the spacelike axis.

The inherent limitation of this approach is that it produces only bounds on the quantities of interest, rather
than definite determinations.   Furthermore, when the experimental errors
are large, there is no significant advantage in adding more and
more input information, as the bounds do not improve.  On the other hand,  due to the increased accuracy of the input observed recently, the bounds can be very stringent, competing in precision with experimental data or theoretical predictions. 

 Applications of this formalism to the pion form factor were undertaken in  \cite{IC} and more recently in \cite{Ananthanarayan:2011xt, Ananthanarayan:2012tn, Ananthanarayan:2012tt}.
In \cite{Ananthanarayan:2011xt}, the phase in the elastic region $t < \tin$, where  $\tin$ is the first important inelastic threshold set by the production of an $\omega\pi$ pair, together with the modulus above $\tin$  used in an averaged way  and some spacelike information, were exploited  for deriving constraints on the higher shape parameters
$c$ and $d$ appearing in the expansion (\ref{eq:taylor}). 

In a more recent work \cite{Ananthanarayan:2012tt}, the same input was used for deriving  bounds on the modulus in 
the elastic region, below the inelastic threshold. The results provided nontrivial consistency
checks on the recent experimental data  \cite{BABAR}-\cite{Fujikawa:2008ma}  on the modulus 
from $e^+ e^-$ annihilation and $\tau$-decay experiments. 
In particular, at low energies the calculated bounds offered 
a more precise description of the modulus than the experimental data.

In \cite{Ananthanarayan:2011xt, Ananthanarayan:2012tn, Ananthanarayan:2012tt}, 
the experimental data on the modulus below the inelastic threshold were not included as input.  
The results showed that the knowledge of the phase in the elastic region has strong implications 
for constraining the shape of the form factor below and above the unitarity threshold.  
The inclusion of the experimental measurements  on the modulus of the form factor below the 
inelastic threshold is the object of the present work. More precisely, we ask what is the 
import of these measurements on the radius $\langle r_\pi^2 \rangle$ and the higher derivatives  
$c$ and $d$ at $t=0$.   
Thus, the present work extends our previous analysis of the shape parameters 
in \cite{Ananthanarayan:2011xt}, where the information on the modulus below $\tin$ was not used. 
It complements the work done in Refs. \cite{Ananthanarayan:2011xt, Ananthanarayan:2012tn, Ananthanarayan:2012tt}
and provides a further consistency check on the experimental data.

There is a rich discussion in the literature on the determination of the shape parameters.
On the theoretical side, a fit based on ChPT to two-loop accuracy 
for $\tau$ decays gives $\langle r_\pi^2 \rangle= (0.431\pm 0.020\pm 0.016)\, {\rm fm}^2$
and $c=(3.2 \pm 0.5_{\rm exp} \pm 0.9_{\rm theor})\,\gev^{-4}$  \cite{CFU}. 
The pion form factor has been calculated in two-loop ChPT which gives  
$\langle r_\pi^2 \rangle=(0.437\pm0.016)\,{\rm fm}^2$ and $c=(3.85 \pm 0.60)\,\gev^{-4}$ \cite{BiCo}. In 
\cite{BiTa}, the values $\langle r^2_\pi \rangle=(0.452\pm0.013$)fm$^2$ and 
$c=(4.49 \pm 0.28)\,\gev^{-4}$ were obtained within the
three flavour framework and at next-to-next-to-leading order in ChPT. 
A theoretical determination based on quark Dyson-Schwinger equation
combined with Bethe-Salpeter equation for meson amplitudes yields
 $\langle r^2_\pi \rangle=0.45\, {\rm fm}^2$~\cite{Maris:1999bh}.
Recent lattice calculations with chiral extrapolation based on two-loop ChPT give 
$\langle r_\pi^2 \rangle =0.409(23)(37)\,{\rm fm}^2$ and $c = 3.22(17)(36)\gev^{-4}$ \cite{Aoki}. 
Phase (Omn\`es-type) representations
with various parametrizations of the phase 
along the whole unitarity cut give $\langle r_\pi^2 \rangle=(0.432\pm 0.001)\,{\rm fm}^2$ and 
$c=(3.84 \pm 0.02)\,\gev^{-4}$  \cite{TrYn2}. Recently,
a fit of spacelike data with Pad\'{e} approximants \cite{Masjuan} gave
the values $\langle r^2 \rangle=(0.445\pm 0.002\pm0.007)\,{\rm fm}^2$ and 
$c = (3.30 \pm 0.03 \pm 0.33)$ GeV$^{-4}$.

 Other values are $\langle r_\pi^2 \rangle=(0.427\pm0.010)\,{\rm fm}^2$, quoted in \cite{Amendolia:1986wj}, and  
$\langle r_\pi^2 \rangle^{\frac{1}{2}}=(0.711\pm 0.009)\,{\rm fm}$ given  in \cite{Bebek},
both being obtained by fits of the data with a simple pole. 
The curvature $c$ has also been determined from fits of experimental data with specific analytic 
parametrizations of the form factor. The value $c= (3.90 \pm  0.10)\,\gev^{-4}$  has been obtained in  \cite{Truo} 
by a standard dispersion relation. A fit of the ALEPH data \cite{Aleph} on the hadronic $\tau$ decay rate 
with a Gounaris-Sakurai formula for the form factor \cite{GoSa}  gives  $c= (3.2 \pm  1.0)$ GeV$^{-4}$.
The analysis based on the  technique of unitarity bounds \cite{Ananthanarayan:2011xt} mentioned above
led to the range $3.75 \gev^{-4}\lesssim c \lesssim 3.98\gev^{-4}$.

The next shape parameter $d$ is much less well known.
Theoretical results from ChPT and  lattice calculations 
are not yet available. The value $d=(9.70\pm 0.40) \, {\rm GeV}^{-6}$ 
has been obtained  from fits of data by means of usual dispersion relations \cite{Truo}, 
while the Taylor expansion of the  Gounaris-Sakurai parametrization \cite{Aleph}, 
leads to $d= 9.80 \, {\rm GeV}^{-6}$. In \cite{Ananthanarayan:2011xt} we derived the range
$9.91 \gev^{-6}\lesssim d \lesssim 10.46\gev^{-6}$.

Except for the ranges predicted in \cite{Ananthanarayan:2011xt},  the previous determinations of the shape parameters came from specific parametrizations, like polynomial expansions, Pad\'e approximants or 
the Gounaris-Sakurai model for the dominant $\rho$ pole. In contrast, the predictions of the present work are parametrization-free in the sense that 
we do not rely on specific analytic expressions of the form factor or its modulus.
The determination follows from general principles of analyticity, with an 
input consisting from the phase in the elastic region and a conservative condition on the  modulus on the unitarity cut. 
The method also avoids the instability problems inherent in the analytic continuation of specific parametrizations.

The scheme of the paper is as follows:  in Sec. \ref{sec:method},
we present very briefly the mathematical formalism whereas in Sec. \ref{sec:inputs}
we describe the  information used as input to our work. In Sec. \ref{sec:results},  
we present the  results of our investigations. In  subsection \ref{subsec:rsq}  we present the analysis of the charge radius  $\langle r_\pi^2 \rangle$ and
in subsection \ref{subsec:cd} the analysis of the higher shape parameters $c$ and $d$. 
In Sec. \ref{sec:conclusions} we summarize our results and present our conclusions.

\section{Method}\label{sec:method}

We use the Fermi-Watson theorem, which states that
\beq\label{eq:watson}
{\rm Arg} [F(t+i\epsilon)]=\delta_1^1(t), \quad\quad t_+ \le t \le\tin,
\eeq
where $\delta_1^1(t)$ is the phase-shift of the $P$-wave of $\pi\pi$ elastic scattering and 
$\tin$ the first inelastic threshold. In addition, we exploit the recent experimental data on the modulus above $\tin$ and the $1/t$ asymptotic decrease of the form factor predicted by perturbative QCD, by adopting the relation
\beq\label{eq:L2}
\D\frac{1}{\pi} \int_{\tin}^{\infty} dt 
\rho(t) |F(t)|^2 = I,
\eeq
where $\rho(t)$ is a suitable positive-definite weight allowing a reliable calculation of the value of $I$. As in \cite{Ananthanarayan:2011xt, Ananthanarayan:2012tn, Ananthanarayan:2012tt} we consider weights of the form
\beq \label{eq:rhogeneric0}
\rho(t) = \frac{t^\beta}{(t+Q^2)^\gamma}, \quad \quad 
\eeq
where  $Q^2 \ge 0$ and $\beta, \gamma$ are taken in the range $\beta\leq \gamma \leq \beta+2$.

The conditions (\ref{eq:watson}) and (\ref{eq:L2}) can be written in a form that allows the application of the mathematical interpolation theory for analytic functions \cite{Meiman, Duren}. We first exploit (\ref{eq:watson}) by introducing the Omn\`es function 
\beq	\label{eq:omnes}
 \omnes(t) = \exp \left(\D\frac {t} {\pi} \int^{\infty}_{t_+} dt' 
\D\frac{\delta (t^\prime)} {t^\prime (t^\prime -t)}\right),
\eeq
where $\delta(t)=\delta_1^1(t)$   for
$t\le \tin$, and is an arbitrary function, sufficiently  smooth ({\em i.e.,}
Lipschitz continuous) for $t>\tin$.
The crucial observation is that the function $h(t)$, defined by
\beq\label{eq:h}
F(t)= \omnes(t) h(t),
\eeq
is real for $t\le \tin$, {\em i.e.} it is analytic in the $t$-plane cut only along $t>\tin$. 
Furthermore, from (\ref{eq:L2}) it follows that $h(t)$  satisfies the condition 
\beq\label{eq:hL2}
\D\frac{1}{\pi} \int_{\tin}^{\infty} dt\, 
\rho(t) |\omnes(t)|^2 |h(t)|^2 = I.
\eeq
This relation can be written in a canonical form by performing the conformal transformation
\beq\label{eq:ztin}
z\equiv \tilde z(t) = \frac{\sqrt{\tin} - \sqrt {\tin -t}} {\sqrt{\tin} + \sqrt {\tin -t}}\,,
\eeq
which maps the upper (lower) lip of the branch-cut $[\tin, \infty]$ to the
upper (lower) half of the unit circle in the complex $z$-plane, and the cut $t$-plane
onto the interior of the unit circle $|z|< 1$, the real line $[-\infty,0]$ to $[-1,0]$
and $[0,\tin]$ to $[0,1]$. 
We introduce then two outer functions, {\it i.e.} functions analytic and without zeros in 
the unit disk $|z|<1$, defined in terms of their modulus on the boundary, related to  
$\sqrt{\rho(t)\, |{\rm d}t/ {\rm d} \tilde z(t)|}$ 
and  $|\omnes(t)|$,  respectively \cite{IC, Abbas:2010EPJA}.
In particular, for weight functions of the form (\ref{eq:rhogeneric0}), the first outer function $w(z)$  
can be written in an analytic closed form in the $z$-variable as \cite{Abbas:2010EPJA}
\beq\label{eq:outerfinal0}
w(z)= (2\sqrt{t_{\rm in}})^{1+\beta-\gamma}\frac{(1-z)^{1/2}} {(1+z)^{3/2-\gamma+\beta}}\frac{(1+\tilde z(-Q^2))^\gamma}{(1-z \tilde z(-Q^2))^\gamma}.
\eeq 
For the second outer function, denoted as $\omega(z)$,  we use an integral representation in terms 
of its modulus on the cut $t>\tin$, which can be written as \cite{IC, Abbas:2010EPJA}
\beq\label{eq:omega}
 \omega(z) =  \exp \left(\D\frac {\sqrt {\tin - \tilde t(z)}} {\pi} \int^{\infty}_{\tin}  \D\frac {\ln |\omnes(t^\prime)|\, {\rm d}t^\prime}
 {\sqrt {t^\prime - \tin} (t^\prime -\tilde t(z))} \right),
\eeq 
where $\tilde t(z)$ is the inverse of $z = \tilde z(t)$, for $\tilde z(t)$  defined in
(\ref{eq:ztin}).

Further, we define a function $g(z)$ by
\beq\label{eq:gF}
 g(z) = w(z)\, \omega(z) \,F(\tilde t(z)) \,[\omnes(\tilde t(z)) ]^{-1},
\eeq 
such that (\ref{eq:hL2}) can be written in terms of the integral of $|g(z)|^2$ on the boundary ($z={\rm e}^{i\theta}$) as
\beq\label{eq:gI1}
\frac{1}{2 \pi} \int^{2\pi}_{0} {\rm d} \theta |g({\rm e}^{i\theta})|^2 = I.
\eeq
The $L^2$-norm equality (\ref{eq:gI1}) is the input condition  to what is known as the Meiman interpolation problem \cite{Meiman}, which consists of finding the most general rigorous correlations between the values of the function and 
its derivatives inside the unit disk $|z|<1$ consistent with the relation (\ref{eq:gI1}) (for a proof and older 
references see \cite{Abbas:2010EPJA}). 

For instance,  with techniques of complex analysis, one can show that (\ref{eq:gI1}) implies  the following determinantal inequality :
\beq\label{eq:det2}
\left|
\begin{array}{c c c c c c}
\bar{I} & \bar{\xi}_{1} & \bar{\xi}_{2} & \cdots & \bar{\xi}_{N}\\	
	\bar{\xi}_{1} & \D \frac{z^{2K}_{1}}{1-z^{2}_1} & \D
\frac{(z_1z_2)^K}{1-z_1z_2} & \cdots & \D \frac{(z_1z_N)^K}{1-z_1z_N} \\
	\bar{\xi}_{2} & \D \frac{(z_1 z_2)^{K}}{1-z_1 z_2} & 
\D \frac{(z_2)^{2K}}{1-z_2^2} &  \cdots & \D \frac{(z_2z_N)^K}{1-z_2z_N} \\
	\vdots & \vdots & \vdots & \vdots &  \vdots \\
	\bar{\xi}_N & \D \frac{(z_1 z_N)^K}{1-z_1 z_N} & 
\D \frac{(z_2 z_N)^K}{1-z_2 z_N} & \cdots & \D \frac{z_N^{2K}}{1-z_N^2} \\
	\end{array}\right| \ge 0,
\eeq  
where the auxiliary quantities
\beq
\bar{I} = I - \sum_{k = 0}^{K-1} g_k^2, \quad  \bar{\xi}_n = g(z_n) - \sum_{k=0}^{K-1}g_k z_n^k\eeq
 are defined in terms of the values :
\bea\label{eq:values} 
\left[\D \frac{1}{k!} \frac{ d^{k}g(z)}{dz^k}\right]_{z=0}&=& g_k, \quad
0\leq k\leq K-1, \nonumber\\
 g(z_n)&=&\xi_n , \quad  \quad1\leq n \leq N.\eea
For simplicity we considered $N$ real points $z_n \in (-1, 1)$  and $K-1$ derivatives at $z=0$. 
The details of the derivation are reviewed in Ref. \cite{Abbas:2010EPJA}.

By using (\ref{eq:gF}) one can express the inequality (\ref{eq:det2}) as a  quadratic constraint on the values of the form factor $F(t)$ and its derivatives at specific points. From this, by solving simple quadratic equations, one can derive upper and lower bounds on one of the first $K$ derivatives at $t=0$, in terms of the other values included in the set.
In particular, if the point $t_n$ is situated on the elastic part of the cut, {\em i.e.} $t_+<t_n<\tin$, the relation (\ref{eq:gFn}) writes as 
\beq\label{eq:gFn}
 g(z_n) = w(z_n)\, \omega(z_n) \,|F(t_n)| /|\omnes(t_n)|,
\eeq 
where $z_n=\tilde z(t_n)$ and the modulus $|\omnes(t)|$ of the Omn\`es function is obtained from (\ref{eq:omnes}) by the Principal Value (PV) Cauchy integral
\beq	\label{eq:modomnes}
 |\omnes(t)| = \exp \left(\frac {t} {\pi} \text{\rm PV} \int^{\infty}_{4 M_\pi^2} dt' 
\D\frac{\delta (t^\prime)} {t^\prime (t^\prime -t)}\right).
\eeq

\section{Input}\label{sec:inputs}
As mentioned above, the first inelastic threshold $\tin$ for the pion form factor is due to the 
opening of the $\omega\pi$ channel which corresponds to $\sqrt{\tin}=M_\omega+M_\pi=0.917\,\gev$. 
Below $\tin$  Fermi-Watson theorem (\ref{eq:watson}) relates the phase of the form factor to the $P$-wave phase shift  
$\delta_1^1(t)$ determined recently with high precision from  Roy equations satisfied
by the $\pi\pi$ elastic amplitude  \cite{ACGL}-\cite{GarciaMartin:2011cn}. 
In fact, some input to these equations 
is borrowed  from  
the knowledge of the pion form factor itself: for instance, the value of the $P$-wave phase shift at the matching 
point of 0.8 GeV  was taken in \cite{ACGL,CGL,Caprini:2011ky} from a Gounaris-Sakurai parametrization of the CLEO 
data on the form factor \cite{Anderson:1999ui}.  As discussed in detail in  Ref. \cite{Ananthanarayan:2012tt},   
this input  may be improved with the new data on the modulus of the form factor.  The purpose of our analyses,  reported in \cite{Ananthanarayan:2011xt, Ananthanarayan:2012tn, Ananthanarayan:2012tt} and continued here,
is precisely to provide analyticity tests leading gradually to a
more accurate determination of the form factor at low energies.

We calculate the Omn\`es function  (\ref{eq:omnes}) using as input for $t\le \tin$ the  phase shift $\delta_1^1(t)$ from Refs. \cite{Caprini:2011ky} and \cite{GarciaMartin:2011cn}, 
which we denote as Bern and Madrid  phase, respectively.
Above $\tin$  we use  a continuous function $\delta(t)$, 
which approaches asymptotically $\pi$. As shown in \cite{Abbas:2010EPJA}, if 
this function is Lipschitz continuous, the dependence on $\delta(t)$ of the functions $\omnes(t)$ and $\omega(z)$,  
defined in (\ref{eq:omnes}) and (\ref{eq:omega}), respectively, exactly compensate  each other, leading to 
results fully independent of the unknown phase in the inelastic region.  This
is one of the important strengths of the method applied in this work.

In our analysis we use the condition (\ref{eq:L2})  with weights of the type (\ref{eq:rhogeneric0}). 
We calculate the integral defined in (\ref{eq:L2}) using the BaBar data \cite{BABAR} 
from $\tin$ up to $\sqrt{t}=3\, \gev$,  continued with  a constant value for the modulus 
in the range $3\, \gev \leq \sqrt{t} \leq 20 \gev$,  smoothly connected with a $1/t$ decrease above 20 GeV. As discussed in \cite{Ananthanarayan:2012tt}, the choice of the weight should  lead to an accurate value for the integral $I$,  providing at the same time a strong constraint on the high energy behaviour of the form factor.
We have tested a large class of parameters in (\ref{eq:rhogeneric0}), and found  that the results for the shape parameters  obtained with various weights are very similar. In our analysis we adopted the weight $\rho(t)=1/t$, for which the value of $I$ is \cite{Ananthanarayan:2012tt}:
\beq\label{eq:Ivalue1}
I=0.578 \pm 0.022, \eeq where the uncertainty is  due to the BaBar experimental errors. 
As shown in \cite{IC,Abbas:2010EPJA}, for a fixed weight the  bounds depend in a monotonous way 
on the value of the quantity $I$, becoming stronger/weaker when this 
value is decreased/increased.  In the applications, we have used as input the central value of $I$ given in 
(\ref{eq:Ivalue1}) increased by the error, which leads to the most conservative bounds.

We use as input also the values at several real points inside the analyticity 
domain $|z|<1$, which translate as $t<\tin$ in the $t$-plane.  Specifically, we implement the normalization condition implemented in the expansion (\ref{eq:taylor}):
\beq\label{eq:norm}
F(0)=1,
\eeq
 and one of the spacelike data taken from \cite{Horn, Huber}:
\bea\label{eq:Huber}	
F(-1.60\,\gev^2)= 0.243 \pm  0.012_{-0.008}^{+0.019}, \nonumber \\ 
F(-2.45\, \gev^2)=  0.167 \pm 0.010_{-0.007}^{+0.013}.
\eea
In addition, we have used the value of the modulus $F(t_n)$ at an energy below  the $\omega\pi$ inelastic threshold,
\beq\label{eq:mod}
|F(t_n)|= F_n \pm \epsilon_n, \quad\quad t_+<t_n<\tin.
\eeq
 with the central value $F_n$ and the error  $\epsilon_n$ taken from one of the recent experiments
\cite{BABAR}-\cite{Fujikawa:2008ma}. 

In our work we implemented the modulus at a single energy and varied this energy over the available experimental range.  Of course, the formalism allows the simultaneous inclusion of more points below the inelastic threshold, and in principle the inclusion of each new input leads to stronger constraints.   However, we recall that the information used as input at the initial points leads to an allowed domain for the value at each new point. On the other hand, each new input is known within an error channel. It may happen that the allowed and the experimental ranges are disjoint. This would signal an inconsistency in the input data, in which case the mathematical problem with an additional input has no solution. If the experimental range of the additional input is consistent with the allowed range determined by the previous values and is narrower than this  range, its inclusion clearly improves the results. However, it turns out that for nonzero experimental errors of the present size, one reaches rather 
quickly a saturation, {\em i.e.}, adding one more datum leads to bounds that are no better than the bounds obtained without that additional datum. 
 In practice, as discussed in \cite{Ananthanarayan:2011xt},   one stops at a reasonable number $N$ when the gain from including a larger number of points is not significant. Our analysis shows that already with one point taken from the spacelike region and another from the timelike region we obtain rather narrow allowed ranges for the quantities of interest.  Thus we stop at $N=2$ and vary the energy of the input modulus. The {\it a posteriori} comparison of the bounds obtained with modulus from different energies will show that actually there are some inconsistencies between different data. 

We mention that we work in the exact isospin limit, taking into account the isospin violation due to the $\rho-\omega$ mixing in $e^+e^-$ annihilation by a standard correction \cite{Leutwyler:2002hm, Hanhart:2012wi}. More exactly, we have divided the experimental  modulus from BaBar, KLOE and CMD-2 experiments by the factor
$|F_{\omega}(t)|$, where
\begin{equation} \label{eq:iso}
F_{\omega}(t)=\Big(1+\epsilon\,\frac{t}{t_\omega-t} \Big), ~~ t_\omega=(M_\omega-i\Gamma_\omega/2)^2, 
\end{equation}
with  $M_\omega=0.7826\,\gev$, $\Gamma_\omega=0.0085\,\gev$ and  $\epsilon=1.9\times 10^{-3}$ \cite{Leutwyler:2002hm, Hanhart:2012wi}. 

An important remark is that, except for the normalization condition (\ref{eq:norm}) which is exact, all the inputs that we use are known only with some uncertainty.  Following the discussion given in \cite{Ananthanarayan:2012tt}, in our analysis we have varied all the inputs simultaneously within their error intervals, taking the most conservative bounds on the quantity of interest  consistent with the input, {\em i.e.} the largest upper bound and the smallest lower bound from the set of values obtained with specific inputs. 
In other words, we take the union of the allowed domains,  obtained with specific input values inside the error intervals, for the quantity of interest. The procedure has been carried out efficiently with a combination of Mathematica and Fortran programs.

\section{Extraction of results}\label{sec:results}

\subsection{$\langle r_\pi^2\rangle$ analysis}\label{subsec:rsq} 
We study first the impact of the conditions (\ref{eq:watson}), (\ref{eq:L2}), (\ref{eq:norm}),  
(\ref{eq:Huber}) and (\ref{eq:mod}) used as input in our formalism  on the charge radius $\langle r_\pi^2\rangle$. 
To this end we calculate upper and lower bounds on the first derivative  appearing in the expansion (\ref{eq:taylor}), 
by applying the general inequality (\ref{eq:det2}) in the particular case $K=2$ and $N=2$. More precisely, we use 
as input, besides the normalization  (\ref{eq:norm}), one of the spacelike values (\ref{eq:Huber}) and a single 
experimental modulus $|F(t_n)|$ at an energy squared $t_n$ below the $\omega\pi$ threshold, from the sets reported in
\cite{BABAR}-\cite{Fujikawa:2008ma}.  The errors on the phase, the spacelike value and the modulus were taken  
into account as explained at the end of Sec. \ref{sec:inputs}. 

\begin{figure}[htb]
\vspace{0.35cm}
\includegraphics[width = 8.cm]{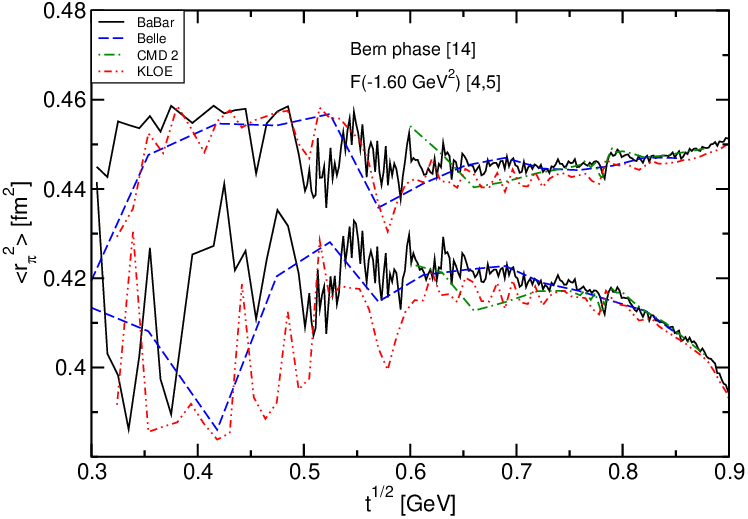}
\caption{Upper and lower bounds on $\langle r_\pi^2\rangle$ using as input one modulus value below the  $\omega\pi$ inelastic threshold from the experiments BaBar, Belle,  CMD-2 and KLOE, as functions of the energy $\sqrt{t}$ where the modulus was implemented. We  used the  Bern phase shift  from \cite{Caprini:2011ky}   and the spacelike value  $F(-1.60\, {\rm GeV}^2)$ given in (\ref{eq:Huber}). }
\label{fig:fig1}
\end{figure}

\begin{figure}[htb]
\vspace{0.35cm}
\includegraphics[width = 8.cm]{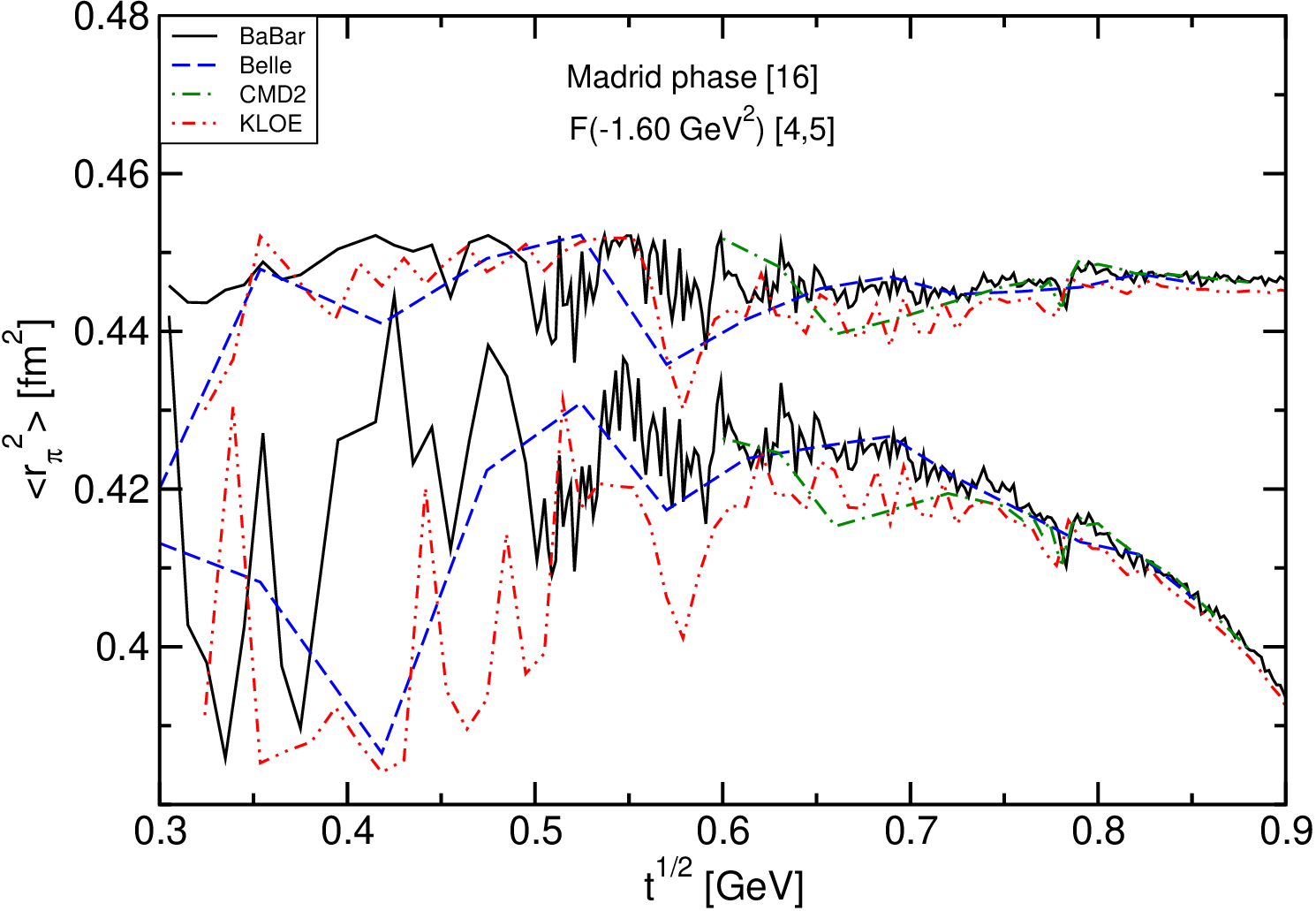}
\caption{As in Fig. \ref{fig:fig1} using as input the Madrid phase from   \cite{GarciaMartin:2011cn}. }
\label{fig:fig2}
\end{figure}

In Figs. \ref{fig:fig1} and \ref{fig:fig2} we present  upper and lower bounds on  $\langle r_\pi^2\rangle$ as functions of the 
energy $\sqrt{t}$ corresponding to the input modulus, for the input phase from  \cite{Caprini:2011ky} and \cite{GarciaMartin:2011cn}, respectively.  We have used input data from BaBar \cite{BABAR}, KLOE 
\cite{KLOE1, KLOE2},  CMD-2 \cite{CMD2:1,CMD2:2} and Belle \cite{Fujikawa:2008ma}
along the whole elastic region from the threshold  $4M_\pi^2$ to the first inelastic threshold $\tin$. For 
convenience we have connected the individual points by  continuous lines.\footnote{The number  of experimental 
points below the $\omega\pi$ threshold reported by the BaBar experiment \cite{BABAR} is 221, while KLOE 
\cite{KLOE1, KLOE2} measured the modulus at 121 energies below $\tin$, CMD-2 \cite{CMD2:1,CMD2:2} at 34 
and Belle \cite{Fujikawa:2008ma} at 15 energies.}  We show the results obtained with the input from the first spacelike point  $t=-1.60\,\gev^2$. The second spacelike point from (\ref{eq:Huber}) gives similar results.

  The curves in Figs. \ref{fig:fig1} and \ref{fig:fig2}  are not smooth: 
the bounds obtained for $\langle r_\pi^2\rangle$ vary quite drastically from point to point, especially at low energies,
which means that they depend strongly on the experimental modulus used as input. 
In addition, there are several experimental data for which we do not even obtain real bounds 
for $\langle r^2_\pi\rangle$: the quadratic equations derived from the inequality (\ref{eq:det2}) 
have complex solutions for all the input values inside the error bars. In these cases (not shown 
in Figs. \ref{fig:fig1} and \ref{fig:fig2})  the input modulus does not satisfy the analyticity constraints with the 
phase and the spacelike values used as input. Such inconsistencies in the data on modulus were 
noted already in our prior analysis \cite{Ananthanarayan:2012tt}.

The upper and lower bounds shown in Figs. \ref{fig:fig1} and  \ref{fig:fig2} define, for each input modulus at a fixed energy, an allowed interval for the charge radius  $\langle r_\pi^2\rangle$, which turns out to be quite narrow for  some particular energies.  In principle, if the data were consistent among each other, the final allowed domain for  $\langle r_\pi^2\rangle$ would be given by the intersection of the  allowed ranges obtained with particular inputs at fixed energies.  The range defined by this intersection will be limited by the smallest upper bound and the largest lower bound shown in the figure. Of course, if the smallest upper bound turns out to be smaller than the largest  lower bound the  intersection is empty. In practice this turns out to be the case,  
if we consider all the points shown in Figs. \ref{fig:fig1} and  \ref{fig:fig2}. This is seen in the left part of Table \ref{table:Int}, where $\langle r_\pi^2\rangle_{\rm max }$ is the lowest upper bound and $\langle r_\pi^2\rangle_{\rm min }$ is the largest lower bound upon all the experimental points below the $\omega\pi$ threshold. The corresponding values for the four experiments are indicated separately. For all experiments $\langle r_\pi^2\rangle_{\rm max }< \langle r_\pi^2\rangle_{\rm min }$, which clearly indicates that there are inconsistencies between the measurements of the modulus at different energies, if we consider all the data between the elastic threshold $t_+$ and the inelastic threshold $\tin$. The extraction of  $\langle r_\pi^2\rangle$ is possible only if we restrict the intersection to a smaller energy interval.

 It is obvious that the low energy region is questionable. The fairly drastic variations of the bounds  obtained with the input from these energies  may be explained by the big fluctuations of the data in this region and their  mutual inconsistencies in spite of the large errors. Therefore, the data on modulus from the low energy region do not allow a consistent extraction of the radius. 

\begin{figure}[htb]
\vspace{0.45cm}
 \includegraphics[width = 8.cm]{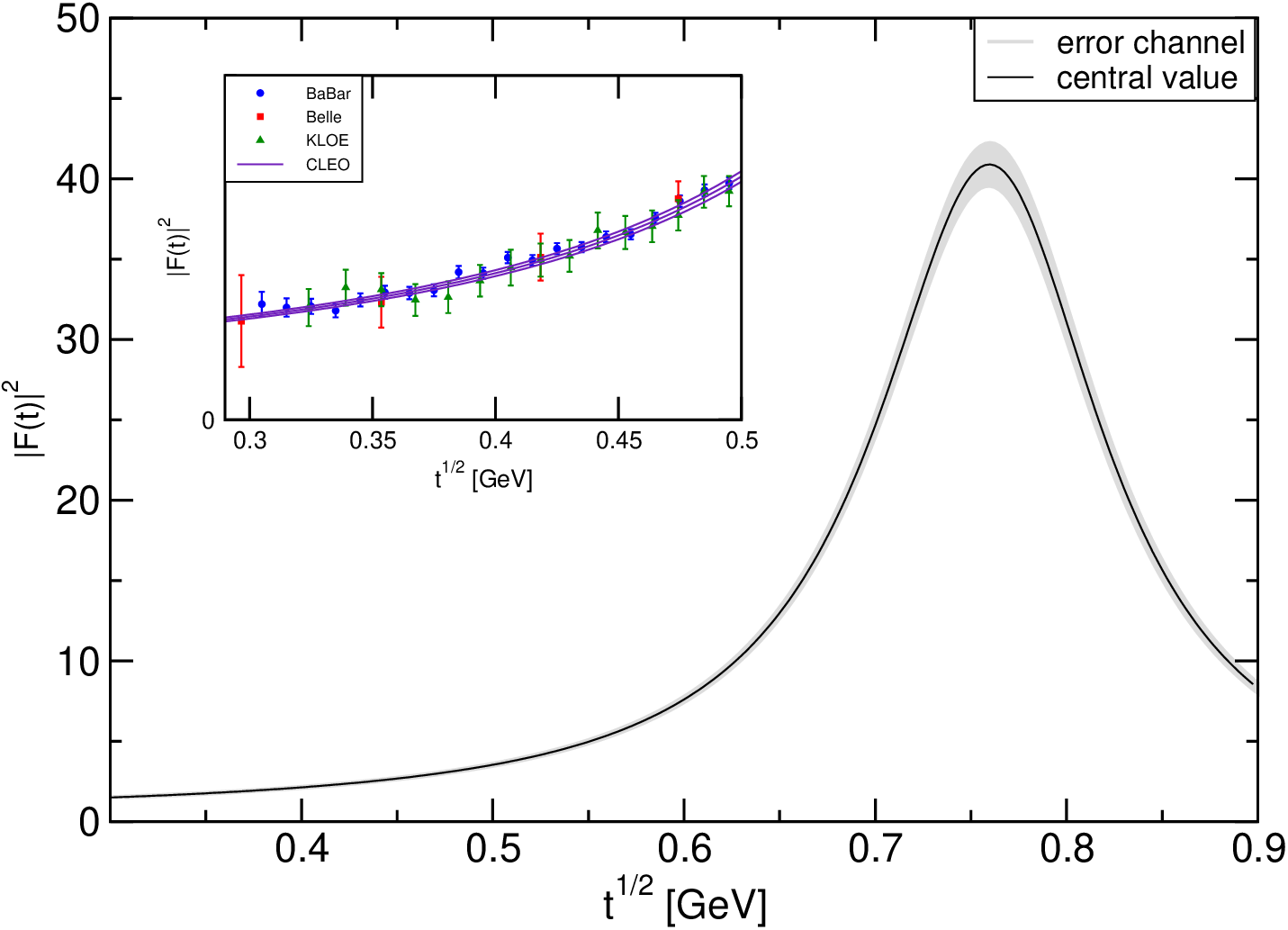}
\caption{Gounaris-Sakurai  parametrization of the CLEO data  \cite{Anderson:1999ui} on the form factor  modulus. The inset shows the parametrization together with the recent data below 0.5 GeV. }
\label{fig:fig3}
\end{figure}

\begin{figure}[htb]
\vspace{0.45cm}
 \includegraphics[width = 8.cm]{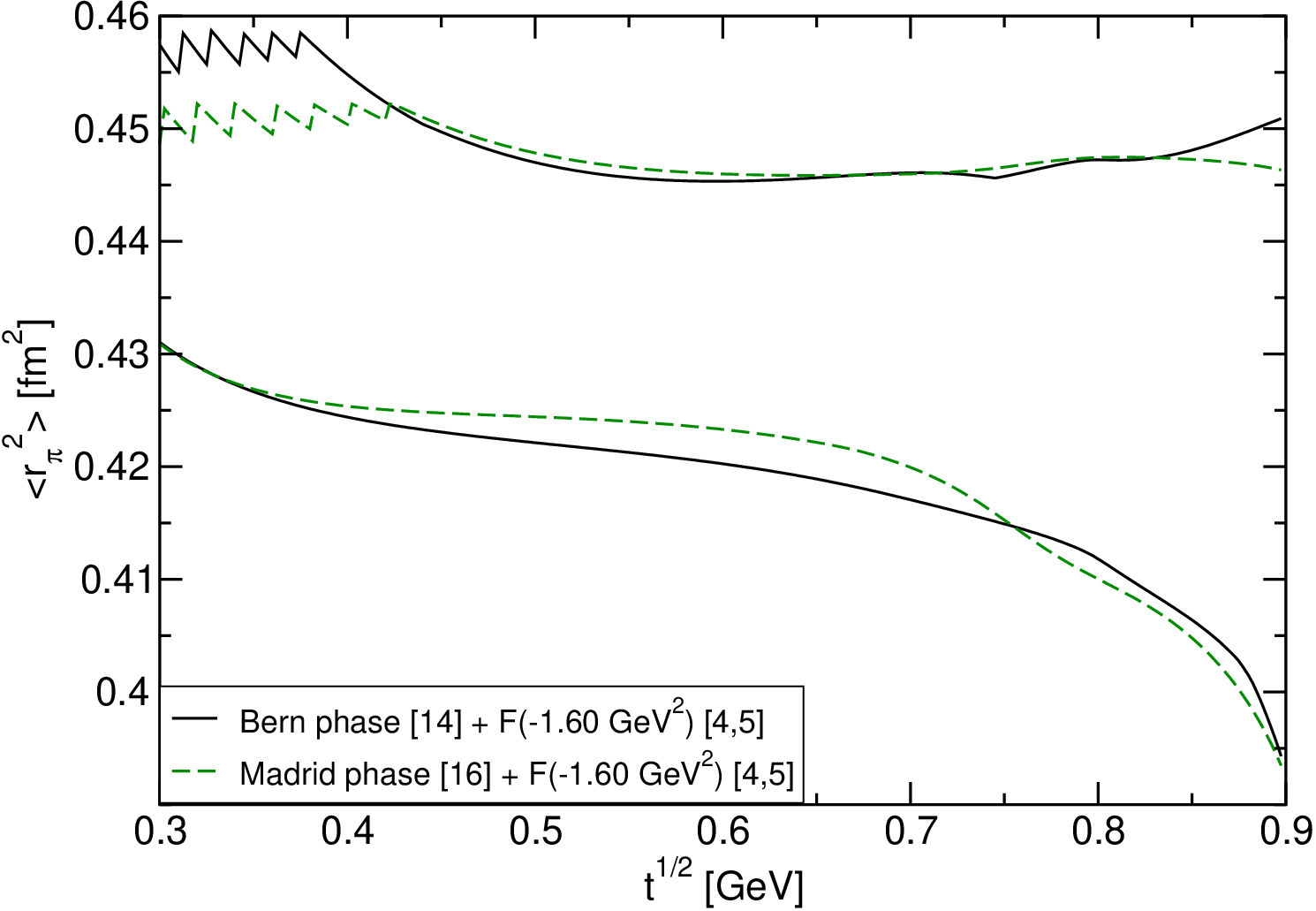}
\caption{Bounds on the charge radius using as input the modulus from the Gounaris-Sakurai  parametrization of the CLEO data  \cite{Anderson:1999ui}.  }
\label{fig:fig4}
\end{figure}

In order to test our formalism with a smoother input, 
we have considered the Gounaris-Sakurai (GS) based parametrization
of the CLEO data  on the modulus \cite{Anderson:1999ui}, shown in Fig. \ref{fig:fig3}.  In fact, the fit adopted in \cite{Anderson:1999ui}
does not satisfy  all the constraints that the form factor must obey (in particular, as shown in \cite{Anderson:1999ui}, the normalization of the fit is allowed to float, and the resulting
best fit explicitly violates the condition (\ref{eq:norm}), being slightly larger than unity at $t=0$). We note also that only data above 0.5 GeV were included in the fit performed in \cite{Anderson:1999ui}, so the curve shown in Fig.  \ref{fig:fig3} at low energies is actually an extrapolation. In our analysis we use the parametrization only to generate input on the unitarity cut, and  impose the normalization condition on the  admissible class of functions used in the derivation of the bounds.

To simulate the experimental situation, we have generated the modulus  at 121 discrete points, using them and the corresponding errors\footnote{For simplicity, we assumed that the errors of the parameters are not correlated.}
as input in our formalism. The upper and lower bounds   on $\langle r_\pi^2\rangle$
obtained in this way for the two phases \cite{Caprini:2011ky} and \cite{GarciaMartin:2011cn}
 are shown in Fig. \ref{fig:fig4}  as functions of the energy at which the input modulus was used. 
One may see that the bounds depend now smoothly  on the input energy, except for a few very small fluctuations at low energies, which may be explained by the fact that the simulated data do not satisfy the normalization condition $F(0)=1$ imposed in the formalism.  The normalization chosen in \cite{Anderson:1999ui}, which artificialy increases the input modulus at low energies, can explain also the higher values of both the upper and lower bounds at the left edge of the figure.
 However, for energies above 0.5 GeV where the fit is based on experimental data,  the 
bounds  are consistent with those given in Figs. \ref{fig:fig1} and \ref{fig:fig2},  except for the fact that the curves are now smooth. This indicates that our formalism is sensitive to the noise in the input, as seen  in Figs. \ref{fig:fig1} and \ref{fig:fig2}, where we have not resorted to  parametrizations of the data. 

\begin{figure}[htb]
\vspace{0.35cm}
\includegraphics[width = 7.8cm]{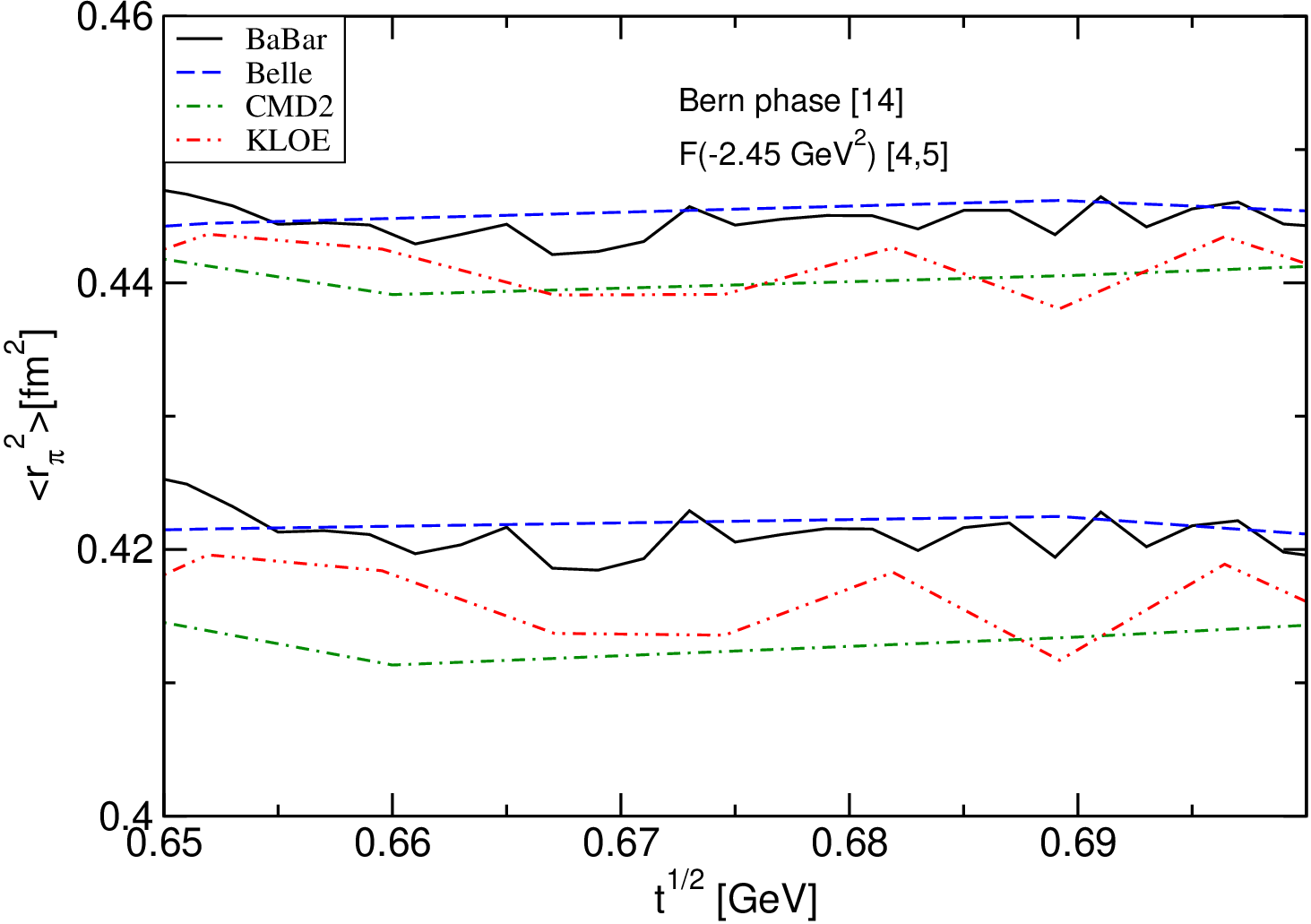}
\caption{Upper and lower bounds on  $\langle r_\pi^2\rangle$ as a function of the energy in the region of stability where the modulus was implemented. }
\label{fig:fig5}
\end{figure}

\begin{figure}[htb]
\vspace{0.35cm}
\includegraphics[width = 7.8cm]{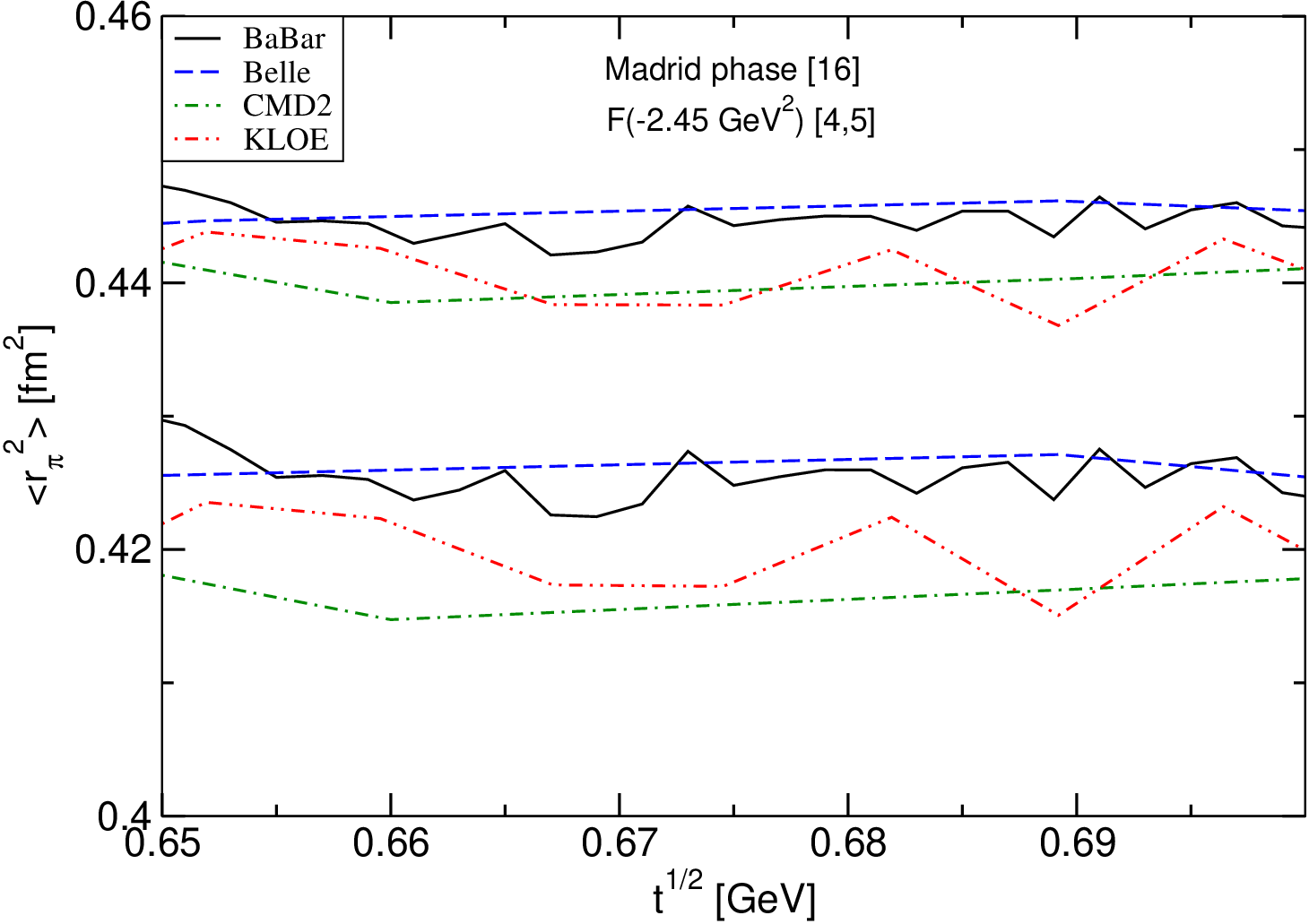}
\caption{As in Fig. \ref{fig:fig5}, using the input phase from \cite{GarciaMartin:2011cn}. }
\label{fig:fig6}
\end{figure}

\begin{center}
\begin{table*}[htpb]
\caption{Intersections of the ranges of $\langle r_\pi^2\rangle$ obtained with various inputs.}\vspace{0.1cm}
\label{table:Int}
\begin{tabular}{|c|c|c|c|c|c|c|c|c|c|} \hline
\multirow{3}{*}{Spacelike data}&\multirow{3}{*}{$|F(t)|$}&\multicolumn{4}{|c|}
{Bern phase \cite{Caprini:2011ky}}&\multicolumn{4}{|c|}{Madrid phase \cite{GarciaMartin:2011cn}}\\
\cline{3-10}
&&\multicolumn{2}{|c|}{All points included}&\multicolumn{2}{|c|}
{(0.65 -- 0.70) GeV}&\multicolumn{2}{|c|}{All points included}&\multicolumn{2}{|c|}{(0.65 -- 0.70) GeV}\\
\cline{3-10}
&&$\langle r_\pi^2\rangle_{\rm min}$ &$ \langle r_\pi^2\rangle_{\rm max} $&$\langle r_\pi^2\rangle_{\rm min}$&$\langle r_\pi^2\rangle
_{\rm max}$&$\langle r_\pi^2\rangle_{\rm min}$&$\langle r_\pi^2\rangle_{\rm max }$&$\langle r_\pi^2\rangle_{\rm min}$&$\langle r_\pi^2\rangle_{\rm max}$\\
\hline
\multirow{4}{*}{$F(-1.60 {\rm GeV}^2$)}&Belle &0.4281&0.4177&0.4227&0.4446&0.4309&0.4184&0.4266&0.4447\\
\cline{2-10}
&BaBar&0.4415&0.4358&0.4234&0.4431&0.4444&0.4361&0.4295&0.4430\\
\cline{2-10}
&CMD2&0.4232&0.4404&0.4171&0.4404&0.4264&0.4396&0.4194&0.4396\\
\cline{2-10}
&KLOE&0.4304&0.4293&0.4201&0.4394&0.4312&0.4300&0.4235&0.4381\\
\cline{2-10}

\hline\hline
\multirow{5}{*}{$F(-2.45 {\rm GeV}^2)$}
&Belle &0.4283&0.4173&0.4225&0.4436&0.4314&0.4179&0.4271&0.4437\\
\cline{2-10}
&BaBar&0.4418&0.4349&0.4232&0.4421&0.4444&0.4351&0.4301&0.4421\\
\cline{2-10}
&CMD2&0.4230&0.4391&0.4161&0.4391&0.4267&0.4385&0.4193&0.4385\\
\cline{2-10}
&KLOE&0.4305&0.4289&0.4196&0.4381&0.4317&0.4288&0.4235&0.4368\\
\hline
\end{tabular}
\end{table*}
\end{center}

\begin{table*}[htpb]
\centering
\caption{Weighted average of upper and lower bounds for $\langle r_\pi^2\rangle$ obtained with various inputs.}\vspace{0.1cm}
\label{table:Ia}
\begin{tabular}{|c|c|c|c|c|c|}\hline
\multirow{2}{*}{Spacelike data}&\multirow{1}{*}{$|F(t)|$}&\multicolumn{2}{|c|}
{Bern phase \cite{Caprini:2011ky}}&\multicolumn{2}{|c|}{Madrid phase \cite{GarciaMartin:2011cn}}\\
\cline{3-6}
&&$\langle r_\pi^2\rangle_{\rm min}$&$\langle r_\pi^2\rangle_{\rm max}$&$\langle r_\pi^2\rangle_{\rm min}$&$\langle r_\pi^2\rangle_{\rm max}$\\
\hline
\multirow{4}{*}{$F(-1.60 {\rm GeV}^2$)}&Belle &0.4152&0.4455&0.4162&0.4437\\
\cline{2-6}
&BaBar&0.4172&0.4470&0.4187&0.4464\\
\cline{2-6}
&CMD2&0.4136&0.4457&0.4125&0.4457\\
\cline{2-6}
&KLOE&0.4045&0.4467&0.4041&0.4446\\
\cline{2-6}

\hline\hline
\multirow{4}{*}{$F(-2.45 {\rm GeV}^2)$}&Belle &0.4136&0.4432&0.4151&0.4426\\
\cline{2-6}
&BaBar&0.4159&0.4463&0.4180&0.4455\\
\cline{2-6}
&CMD2&0.4117&0.4447&0.4106&0.4447\\
\cline{2-6}
&KLOE&0.4005&0.4456&0.4005&0.4434\\
\hline
\end{tabular}
\end{table*}

 Taking into account the discussion above, for the real data we restrict our calculations to the input from energies above 0.65 GeV, which is free of the big fluctuations observed at low energies. We note that the bounds show a remarkable stability in the region (0.65-0.70) GeV, and become weaker for the input from energies above 0.70 GeV.  The upper and lower bounds on $\langle r_\pi^2\rangle$ obtained with various
data sets  in the region of stability  as functions of the input energy are compared  in Figs. \ref{fig:fig4}  and \ref{fig:fig5}.  We show for illustration the bounds obtained with the second spacelike point from (\ref{eq:Huber}), which are slightly better than those obtained with the first point due to its smaller relative error.   

  One can notice that the bounds obtained with the Madrid phase  \cite{GarciaMartin:2011cn} are 
slightly shifted upwards compared
to those obtained with the Bern phase \cite{Caprini:2011ky}. 
We also see that the bounds from BaBar \cite{BABAR} and Belle \cite{Fujikawa:2008ma} 
are shifted upwards compared to the  bounds obtained with the  data of CMD-2 \cite{CMD2:2} and KLOE \cite{KLOE1,KLOE2}.

If we restrict the input to energies above 0.65 GeV, we obtain a nonzero  intersection of the allowed ranges for the charge radius obtained with individual data on the modulus.  Actually, the energies above 0.7 GeV lead to weaker  bounds and do not influence the intersection. So, we can use in the numerical calculations only the stable region (0.65-0.70) GeV.  We show the results of the intersection in  Table \ref{table:Int}, where, for each data set,  $\langle r_\pi^2\rangle_{\rm max }$ is the lowest upper bound and $\langle r_\pi^2\rangle_{\rm min }$ is the largest lower bound when the energy of the input modulus is varied along the stable region.  
From the values listed in the Table for all data sets we infer that the final range obtained by the intersection of the particular ranges is given by 
\beq\label{eq:inters}
\langle r_\pi^2\rangle_{\rm min }\approx 0.42\,{\rm fm}^2,\quad \langle r_\pi^2\rangle_{\rm max }\approx 0.44\,{\rm fm}^2.
\eeq

Since a strict intersection of the $\langle r_\pi^2\rangle$  ranges upon the whole energy region
does not exist, we may adopt a weaker definition of the final allowed range, 
by taking the weighted average of the upper (lower) bounds on $\langle r_\pi^2\rangle$ shown in Fig. \ref{fig:fig1}.
It is reasonable to use as weights in the average the experimental errors $\epsilon_n$ on the
measured modulus $F_n$ as defined in (\ref{eq:mod}). Thus we calculate for both the upper and lower bounds the weighted averages 
\begin{equation}\label{eq:aver}
\langle r_\pi^2\rangle_{av}=\frac{\sum_n w_n \langle r_\pi^2\rangle_n}{\sum_n w_n}, \quad \quad w_n=1/\epsilon_n^2,
\end{equation}
where the sum runs over the individual energy points $t_n$. This definition suppresses the contribution of the low-energy data, which have larger errors.  The results are presented for all data sets in Table \ref{table:Ia}, from which we derive the weaker bounds
\beq\label{eq:inters1}
\langle r_\pi^2\rangle_{\rm min,\,av }\approx 0.40\,{\rm fm}^2,\quad \langle r_\pi^2\rangle_{\rm max,\,av }\approx 0.45\,{\rm fm}^2.
\eeq
The range defined by the limits given in  (\ref{eq:inters}) and the more conservative limits (\ref{eq:inters1}) are our parametrization-free determinations based on the recent data on the modulus.

\subsection{$c$-$d$ analysis}\label{subsec:cd}
We now turn to  the higher shape parameters $c$ and $d$ in the Taylor expansion (\ref{eq:taylor}).
In our previous work  \cite{Ananthanarayan:2011xt} we have obtained an allowed  domain in the $c$-$d$ plane 
using as input the phase in the elastic region, the integral condition (\ref{eq:L2})  and the experimental spacelike data given in  (\ref{eq:Huber}).  The  modulus of the form factor 
in the timelike region below the inelastic threshold was not used as input in that study. Instead, we had used as input a range for the radius  $\langle r_\pi^2 \rangle $ taken from the literature.
In the present work, we update the analysis by including as input the modulus measured below the inelastic threshold.  It turns out that this input leads to strong constraints on the higher shape parameters at $t=0$ without any information on the charge radius. 

The allowed domain in the $c$-$d$ plane is obtained from the inequality (\ref{eq:det2}) written in the particular case  $K=4$,  using as input the normalization (\ref{eq:norm}) and one ($N=1$) or two ($N=2$) additional values at interior points. One can see that the first derivative at $t=0$, {\em i.e.} the radius  $\langle r_\pi^2 \rangle $, appears in the coefficients of the quadratic relation for the parameters  $c$ and $d$ resulting from the determinant (\ref{eq:det2}). Moreover, the quadratic equation for $c$ at a fixed $d$, for instance, has real solutions only for  $\langle r_\pi^2 \rangle $ in the allowed range determined from   the same inequality (\ref{eq:det2}), as shown in the previous subsection.
So, in principle it is not necessary to put a limitation on the range of the charge radius when deriving constraints on the higher shape parameters $c$ and $d$: all we have to do is to vary the value of  $\langle r_\pi^2 \rangle $ over an interval large enough as to cover the allowed range, and take the weakest bounds obtained with this variable input. 

However, if the modulus below $\tin$  is not included, the constraint on the radius is rather weak, and the corresponding domain in the $c$-$d$ plane will be quite large.
Therefore, in this case one has to adopt as input a certain smaller range for   $\langle r_\pi^2 \rangle $ in order to actually constrain the higher derivatives. In \cite{Ananthanarayan:2011xt} we have taken as input the narrow range $\langle r_\pi^2 \rangle = (0.435 \pm 0.005)\,{\rm fm}^2$, suggested by some previous works.  We now repeated first that analysis, using for $\langle r_\pi^2 \rangle $ the larger range  $(0.42-0.44)\, {\rm fm}^2$,  derived in Eq. (\ref{eq:inters}).

\begin{figure}[htb]
\vspace{0.3cm}
 \includegraphics[width = 8.cm]{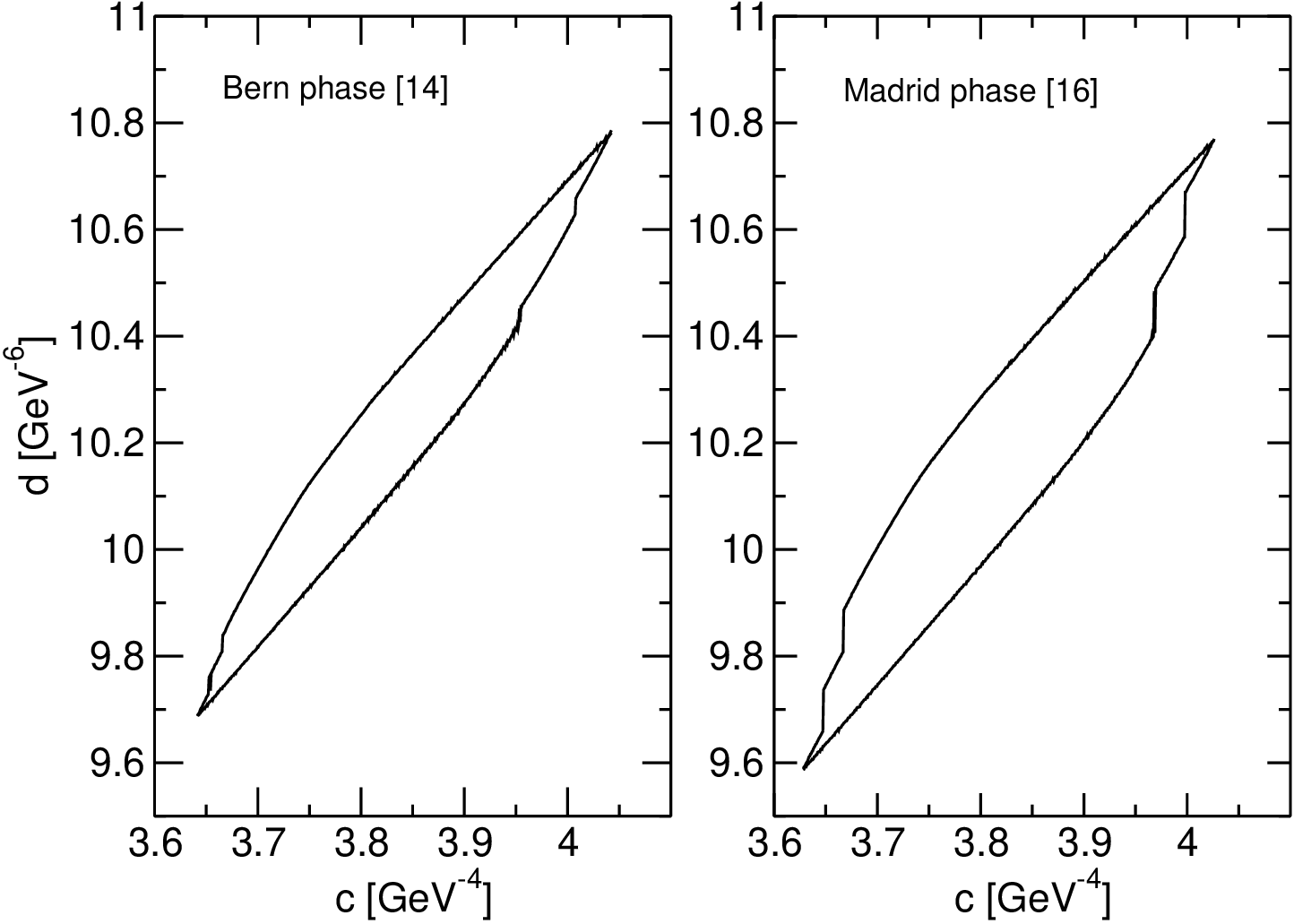}
\caption{Allowed domain in the  $c$-$d$ plane obtained without timelike modulus data.}
\label{fig:fig7}\vspace{0.3cm}
\end{figure}

The results are presented in Fig. \ref{fig:fig7} for the two input phases, 
Bern \cite{Caprini:2011ky} and Madrid \cite{GarciaMartin:2011cn} and the 
spacelike value at -2.45 GeV$^2$.  We recall that for a fixed input the 
allowed domain is the interior of an ellipse in the $c$-$d$ plane, whose 
boundary  is given by the upper and lower bounds on $d$ at each fixed $c$, 
found by solving a quadratic equation.  The errors on the phase and 
spacelike value were taken into account,  as explained at the end of 
Sec.  \ref{sec:inputs}, by varying the input quantities inside the error 
intervals and taking the weakest bounds, {\em i.e.} the union of the 
corresponding ranges of $d$ at each fixed $c$. The final allowed domain is no longer an ellipse, but has a more complicated shape.

 Fig. \ref{fig:fig7} shows that the domains obtained are consistent  with 
the ranges of $c$ and $d$ quoted in \cite{Ananthanarayan:2011xt}, being slightly weaker due to the more conservative input for the charge radius adopted now.  Note that we  use also a different  weight function $\rho(t)$ in the condition (\ref{eq:L2}): while in  \cite{Ananthanarayan:2011xt} we adopted a special weight relevant for the calculation of the muon's $(g-2)$, we now make calculations with  $\rho(t)=1/t$. In fact, we have tried also other weights of the form (\ref{eq:rhogeneric0}), and found that the constraints on the shape parameters at $t=0$ are not very sensitive to the choice of the weight. 

\begin{figure}[htb]
\vspace{0.3cm}
 \includegraphics[width = 8.cm]{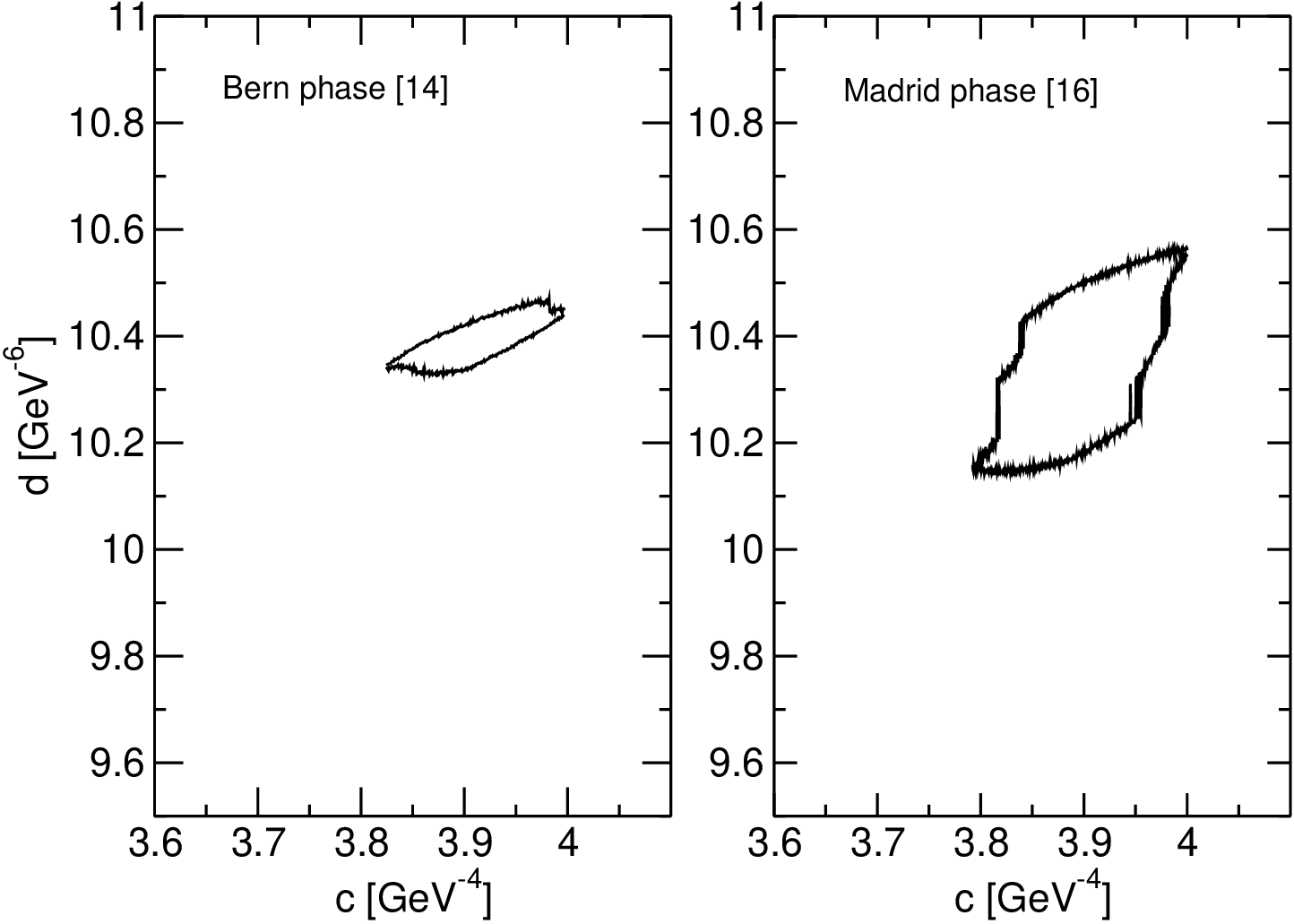}
\caption{Allowed domain in the  $c$-$d$ plane obtained with modulus data from BaBar \cite{BABAR}
in the energy region 0.65-0.70 GeV.}
\label{fig:fig8}\vspace{0.3cm}
\end{figure}

We then include as input the value of the modulus $|F(t_n)|$ at a fixed energy $t_n$ below the inelastic threshold. 
In this analysis we used only BaBar data from the stability region defined above.
As we mentioned, in this case it is not necessary to restrict the charge radius, because the allowed range imposed by the input is already quite narrow. In the numerical calculations, we restricted however
$\langle r^2_\pi\rangle$ to the interval between $0.41\,{\rm fm}^2$ and $0.45\,{\rm fm}^2$, which covers practically
 all the ranges obtained at fixed energies with BaBar data. Extending this range would not have perceptible effects.
 At each energy, we varied also the input quantities (the phase, the spacelike datum and the modulus $|F(t_n)|$) within their error bars,  and took the weakest bounds on $d$ at each fixed $c$.   Finally,  the input energy $t_n$ was varied in the stability region, and the intersection of the particular allowed domains  was derived, by taking the smallest upper bound and the largest lower bound on $d$ at each admissible $c$. The boundary of the resulting allowed domain, which  has a more complicated shape, was found numerically. In Fig. \ref{fig:fig8}  we present the allowed domains in the $c$-$d$ plane obtained by the above procedure.  The improvement brought by the modulus information is 
considerable: the domains are now much  smaller than those in Fig. \ref{fig:fig7}. 
From these domains we predict  the allowed ranges:
\bea\label{eq:cdrange}
&&c \in\, (3.79,\,4.00)\,{\rm GeV}^{-4},\nonumber\\
&&d \in\, (10.14,\,10.56)\,{\rm GeV}^{-6}, \eea
with  a strong correlation between the two parameters. 

We emphasize however that the ranges (\ref{eq:cdrange}) 
are based only on the BaBar data  and should be regarded as provisional.  
Indeed, the small inconsistencies beween the data sets, visible already in 
the extraction of the radius, are expected to be amplified  in the higher 
derivatives. Therefore, precise experimental predictions for the higher 
shape parameters will be possible only when the discrepancies between 
the data sets will be understood and consistent data will be available. 

\section{Discussion and conclusions}\label{sec:conclusions}

Our work was motivated by the fact that the phenomenological knowledge of the pion electromagnetic form factor, which improved considerably in recent years, comes from different sources: the phase below 0.917 GeV is known via Fermi-Watson
theorem from the $P$-wave phase shift of $\pi\pi$ scattering calculated from the Roy equations, the modulus has been measured recently by high statistics experiments on $e^+e^-$ annihilation or $\tau$-decays, and  measurements  on the spacelike axis from electroproduction data  with improved accuracy were also reported.  Suitable tools based on analyticity must be devised for  exploiting the available knowledge in order to make model-independent predictions in regions not directly accessible to experiment. 

 If the modulus and the phase are both  known along an energy interval of the unitarity cut, the function can be reconstructed in principle everywhere, by the uniqueness of analytic continuation. However, since the input is known with some uncertainty, the reconstruction is practically impossible. Indeed,  analytic continuation is unique, but it is also an unstable (ill-posed) problem in the Hadamard sense \cite{Ciulli}, which means that the small uncertainties  along the original range may be amplified arbitrarily in much of the complex plane outside that range. 

In our analysis we applied a formalism that exploits in an optimal way the information on the phase and modulus on the unitarity cut,  leading to results independent on the unknown phase above the inelastic threshold.  
 Combined with the well-known analytic interpolation theory, it allows us to include  information or derive predictions  on the values at points inside the analyticity domain. The formalism also acts as a sensitive devise for testing the consistency of the 
various experimental data sets with analyticity and among themselves. The technique has been already applied to the pion form factor in Refs. \cite{Ananthanarayan:2011xt, Ananthanarayan:2012tn, Ananthanarayan:2012tt}. 

In the present work we studied
the impact of the timelike modulus data below the inelastic 
threshold on the determination of the charge radius $\langle r_\pi^2\rangle$  
and the higher shape parameters $c$ and $d$ in the Taylor expansion (\ref{eq:taylor}). 
To our knowledge, this is the first application of the high-statistics
modulus measurements to the extraction of the pion charge radius
and the higher shape parameters. 

  Using as input
 the conditions (\ref{eq:watson}), (\ref{eq:L2}), (\ref{eq:norm}),  (\ref{eq:Huber}) and (\ref{eq:mod}), 
 we calculated from the inequality (\ref{eq:det2}) upper and lower bounds on the first derivative  appearing in the expansion (\ref{eq:taylor}). The uncertainties of the phase in the elastic region, the spacelike value and the modulus were taken  into account by varying the input inside the error intervals and taking the largest allowed range. The final prediction for  $\langle r_\pi^2\rangle$  is given in principle by the intersection of all the ranges obtained with an input modulus at fixed energy below the inelastic threshold. 

The analysis presented in Sec. \ref{sec:results} shows that, for all the experimental data sets, this intersection turns to be empty, which indicates that there are inconsistencies in the input data. Especially, the data at low energes  lead  to results that strongly fluctuate from point to point and are not compatible among them, in spite of the large experimental errors.  However, if we  restrict the input 
to measurements of the modulus at energies greater than 0.65 GeV,  the results from all the experiments, BaBar, KLOE, CMD-2 and Belle, and the two phases, Madrid and Bern, are consistent to a large extent and lead to the prediction  (\ref{eq:inters}), which we write as
\beq\label{eq:final}
\langle r_\pi^2 \rangle \in\, (0.42,\,0.44)\,\fmsq.  
\eeq
Our parametrization-free prediction (\ref{eq:final})  is consistent with most of the results based on specific parametrizations reported in the literature.

Note that we avoid the presentation of this result in terms of a central value and an error, 
for instance  $\langle r_\pi^2 \rangle = (0.43 \pm 0.01)\,\fmsq$ instead of  (\ref{eq:final}). 
The reason is that, although the probability of the parameters to be in the predicted allowed 
ranges follows from the probabilities of the error intervals inside which we varied the input
 quantities,  we cannot actually attach precise probabilities to the specific values inside the 
allowed domains. Indeed, even for the central values of the input we obtain a range for the 
parameters of interest, not a definite prediction. As a statistical interpretation is difficult 
to give, we can assume that all the values in the predicted range are equally probable.

 For the higher shape parameters $c$ and $d$  the ranges (\ref{eq:cdrange}) 
obtained with the BaBar data are very stringent, illustrating the great 
constraining power of the method. We emphasize however that the  $c$-$d$ 
domains  are very sensitive to the input modulus.  Therefore, 
the results (\ref{eq:cdrange})  should be regarded as provisional until the 
discrepancies between the experimental data sets are understood and consistent 
data are available.

 We recall that in \cite{Ananthanarayan:2012tt} the mathematical
formalism was applied in a somewhat opposite way, {\em i.e.} we adopted
as input the value of the radius and derived bounds on the modulus below
the $\omega\pi$ threshold. In fact, we have used as input precisely the range
(\ref{eq:final}), which we now obtained from the information on the
modulus, and found that the derived bounds are consistent with the
experimental data at higher energies, but in some disagreement at low
energies. Thus, the two analyses are perfectly consistent. In particular,
as already mentioned, the bounds on the modulus at low energies, derived
with the input range   (\ref{eq:final}) for the radius, are more precise
than the data. Therefore,  they can be used for improving the evaluation
of the two-pion contribution to the muon's magnetic anomaly, especially the contribution of the region below 0.5 GeV, which is small but is known to have a relatively large error.  This problem
will be studied in a future work.

\vskip0.3cm

\noindent{\bf Acknowledgement:}  
IC acknowledges support from the Program Nucleu under Contract PN 09370102/2009, and from Program Idei-PCE, Contract No 121/2011.
ISS acknowledges the support from Deutsche Forschungsgemeinschaft
Research Unit FOR 1873 “Quark Flavour Physics and Effective
Theories”, Contract No. KH 205/2-1.


\begin{thebibliography}{100}

\bibitem{Amendolia:1986wj}
  S.R.~Amendolia {\it et al.}  [NA7 Collaboration],
  Nucl.\ Phys.\ B {\bf 277} (1986) 168.

\bibitem{Brown:1973wr}	
C.N. Brown, C.R. Canizares, W.E. Cooper, A.M. Eisner, G.J. Feldmann, C.A. Lichtenstein, L. Litt, W. Loceretz, V.B. Montana and F.M. Pipkin 
  Phys.\ Rev.\  D {\bf 8}, 92 (1973).

\bibitem{Bebek}
  C.J. Bebek {\it et al.},'
  Phys. Rev. D{\bf 17}, 1693 (1978).

\bibitem{Horn}
  T.~Horn {\it et al.}  [Jefferson Lab F(pi)-2 Collaboration],
  Phys.\ Rev.\ Lett.\  {\bf 97} (2006) 192001
  [nucl-ex/0607005].

\bibitem{Huber}
  G.~M.~Huber {\it et al.}  [Jefferson Lab Collaboration],
  Phys.\ Rev.\ C {\bf 78} (2008) 045203
  [arXiv:0809.3052 [nucl-ex]].


\bibitem{BABAR}
  B.~Aubert {\it et al.}  [BABAR Collaboration],
  Phys.\ Rev.\ Lett.\  {\bf 103} (2009) 231801
  [arXiv:0908.3589 [hep-ex]].

\bibitem{KLOE1}
  F.~Ambrosino {\it et al.}  [KLOE Collaboration],
  Phys.\ Lett.\ B {\bf 670} (2009) 285
  [arXiv:0809.3950 [hep-ex]].

\bibitem{KLOE2}
  F.~Ambrosino {\it et al.}  [KLOE Collaboration],
  Phys.\ Lett.\ B {\bf 700} (2011) 102
  [arXiv:1006.5313 [hep-ex]].

\bibitem{CMD2:1}
  R.R.~Akhmetshin {\it et al.}, [CMD-2 Collaboration],
  JETP Lett.\  {\bf 84} (2006) 413
   [Pisma Zh.\ Eksp.\ Teor.\ Fiz.\  {\bf 84} (2006) 491]
  [hep-ex/0610016].

\bibitem{CMD2:2}
  R.~R.~Akhmetshin {\it et al.}  [CMD-2 Collaboration],
  Phys.\ Lett.\ B {\bf 648} (2007) 28
  [hep-ex/0610021].

\bibitem{Fujikawa:2008ma}
  M.~Fujikawa {\it et al.}  [Belle Collaboration],
  Phys.\ Rev.\ D {\bf 78} (2008) 072006
  [arXiv:0805.3773 [hep-ex]].

\bibitem{ACGL}
  B.~Ananthanarayan, G.~Colangelo, J.~Gasser and H.~Leutwyler,
  Phys.\ Rept.\  {\bf 353} (2001) 207
  [hep-ph/0005297].

\bibitem{CGL}
  G.~Colangelo, J.~Gasser and H.~Leutwyler,
  Nucl.\ Phys.\ B {\bf 603} (2001) 125
  [hep-ph/0103088].

\bibitem{Caprini:2011ky}
  I.~Caprini, G.~Colangelo and H.~Leutwyler,
  Eur.\ Phys.\ J.\ C {\bf 72} (2012) 1860
  [arXiv:1111.7160 [hep-ph]].

\bibitem{KPY}
  R.~Kaminski, J.~R.~Pelaez and F.~J.~Yndurain,
  Phys.\ Rev.\ D {\bf 77} (2008) 054015
  [arXiv:0710.1150 [hep-ph]].

\bibitem{GarciaMartin:2011cn}
  R.~Garcia-Martin, R.~Kaminski, J.~R.~Pelaez, J.~Ruiz de Elvira and F.~J.~Yndurain,
  Phys.\ Rev.\ D {\bf 83} (2011) 074004
  [arXiv:1102.2183 [hep-ph]].

\bibitem{Ciulli} S. Ciulli, C. Pomponiu and I. Sabba-Stefanescu, Phys. Rept. {\bf 17} (1975) 133.

\bibitem{Abbas:2010EPJA}
  G.~Abbas, B.~Ananthanarayan, I.~Caprini, I.~Sentitemsu Imsong and S.~Ramanan,
  Eur.\ Phys.\ J.\ A {\bf 45} (2010) 389
  [arXiv:1004.4257 [hep-ph]].


\bibitem{IC}
  I.~Caprini,
  Eur.\ Phys.\ J.\ C {\bf 13} (2000) 471
  [hep-ph/9907227].

\bibitem{Ananthanarayan:2011xt}
  B.~Ananthanarayan, I.~Caprini and I.~S.~Imsong,
  Phys.\ Rev.\ D {\bf 83} (2011) 096002
  [arXiv:1102.3299 [hep-ph]].

\bibitem{Ananthanarayan:2012tn}
  B. Ananthanarayan, I. Caprini and I.S. Imsong, 
Phys. Rev. D {\bf 85} (2012) 096006 [arXiv:1203.5398 [hep-ph]]. 

\bibitem{Ananthanarayan:2012tt}
  B.~Ananthanarayan, I.~Caprini, D.~Das and I.~S.~Imsong,
  Eur.\ Phys.\ J.\ C  {\bf 72} (2012) 2192
  [arXiv:1209.0379 [hep-ph]].

\bibitem{CFU}
  G.~Colangelo, M.~Finkemeier and R.~Urech,
  Phys.\ Rev.\  D {\bf 54}, 4403 (1996)
  [arXiv:hep-ph/9604279].

\bibitem{BiCo}
  J. Bijnens, G. Colangelo and P. Talavera,
  JHEP {\bf 9805}, 014 (1998)
  [arXiv:hep-ph/9805389].

\bibitem{BiTa}
  J.Bijnens and P. Talavera,
  JHEP {\bf 0203}, 046 (2002)
  [arXiv:hep-ph/0203049].

\bibitem{Maris:1999bh}
  P.~Maris and P.~C.~Tandy,
  Phys.\ Rev.\ C {\bf 61} (2000) 045202
  [nucl-th/9910033].


\bibitem{Aoki}
  S.~Aoki {\it et al.}  [JLQCD Collaboration and TWQCD Collaboration],
  Phys.\ Rev.\  D {\bf 80}, 034508 (2009)
  [arXiv:0905.2465 [hep-lat]].

\bibitem{TrYn2} 
J.~F.~de Troconiz and F.~J.~Yndurain,
  Phys.\ Rev.\  D {\bf 71}, 073008 (2005)
  [arXiv:hep-ph/0402285].

\bibitem{Masjuan}
  P.~Masjuan, S.~Peris and J.~J.~Sanz-Cillero,
  Phys.\ Rev.\  D {\bf 78}, 074028 (2008)
  [arXiv:0807.4893 [hep-ph]].


\bibitem{Truo} 
T.~N.~Truong,
  arXiv:hep-ph/9809476.

 \bibitem{Aleph} 
 R.~Barate {\it et al.}  [ALEPH Collaboration],
   Z.\ Phys.\  C {\bf 76}, 15 (1997).
 
 \bibitem{GoSa}
 G.~J.~Gounaris and J.~J.~Sakurai,
   Phys.\ Rev.\ Lett.\  {\bf 21}, 244 (1968).


\bibitem{Meiman} N.N. Meiman, Sov. Phys. JETP. {\bf 17} (1963) 830.

\bibitem{Duren} P. Duren, \emph{ Theory of $H^{\rm p}$ Spaces},  Academic Press, New York, 1970.

\bibitem{Anderson:1999ui}
  S.~Anderson {\it et al.}  [CLEO Collaboration],
  Phys.\ Rev.\ D {\bf 61} (2000) 112002
  [hep-ex/9910046].

\bibitem{Leutwyler:2002hm}
  H. Leutwyler, hep-ph/0212324.

\bibitem{Hanhart:2012wi}
  C.~Hanhart,
 Phys. Lett. B{\bf 715} (2012) 170,  arXiv:1203.6839 [hep-ph].


\end{thebibliography}
\end{document}